\documentstyle[aps,epsf,twocolumn]{revtex}
\begin{document}
\title{New type of conductivity oscillations in quantized films with rough walls}
\author{A. E. Meyerovich, I. V. Ponomarev}
\address{Department of Physics, University of Rhode Island, 2 Lippitt Rd., Kingston\\
RI 02881-0817}
\date{\today}
\address{\mbox{ }}
\address{\parbox{14cm}{\rm \mbox{ }\mbox{ }\\A new type of quantum size effect (QSE)
oscillations is predicted for films with a relatively large correlation
radius of surface inhomogeneities. The effect replaces usual QSE for random
inhomogeneities with Gaussian and exponential power spectra. The
well-pronounced oscillations of conductivity $\sigma $ as a function of
channel width $L$ separate two distinct regions with different indices in
the power-law dependence $\sigma \left( L\right) $. The oscillations are
explained and their positions identified. The effect is reminiscent of
magnetic breakthrough and can simplify observation of QSE in metals.\\}}
\address{\mbox{ }}
\address{\vspace{2mm}\parbox{14cm}{\rm PACS numbers: 72.10.Fk, 73.23.Ad,73.50.Bk}}
\maketitle

%\maketitle
%\pacs{72.10.Fk, 73.23.Ad,73.50.Bk}
%\begin{abstract}
%\end{abstract}
Progress in nanofabrication reignited studies of ultrathin films with
quantum size effect (QSE). QSE describes quantization of motion of particle
across the film, $p_{x}\rightarrow \pi j/L$ (below $\hbar =1$), and leads to
a split of the $3D$\ energy spectrum $\epsilon \left( {\bf p}\right) $ into
a set of minibands $\epsilon _{j}\left( {\bf q}\right) $ (${\bf q}$\ is the $%
2D$ momentum along the film). QSE is routinely observed by various
spectroscopic and STM methods (see, {\it e.g.}, \cite{qse2} and references
therein). QSE also leads to a pronounced saw-like dependence of conductivity 
$\sigma $ on, for example, film thickness $L$. Though only few transport
measurements exhibit QSE directly \cite{qse1}, it has been known for a long
time that the saw-like curves $\sigma \left( L\right) $ should exist for
both bulk \cite{rr6} and surface \cite{r2} scattering.

''Usual'' QSE singularities in $\sigma \left( L\right) $ correspond to
abrupt changes in the number $S={\rm Int}\left( L/\lambda _{F}\pi \right) $
of the occupied minibands $\epsilon _{j}$\ in the points when the film
thickness $L$ becomes equal to $L=k\pi \lambda _{F}$ with integer $k$ (the
Fermi wavelength $\lambda _{F}=1/p_{F}$). The drops in $\sigma \left(
L\right) $ in these points are explained by an opening of $k$ new scattering
channels associated with the scattering-driven transitions to and from the
newly accessible highest miniband $\epsilon _{k}$. The amplitude of these
drops (''saw teeth'') is determined by comparison of the interband
transition probabilities $W_{j\neq j^{\prime }}\left( {\bf q-q}^{\prime
}\right) $ with the intraband scattering $W_{jj}\left( {\bf q-q}^{\prime
}\right) $. When the off-diagonal $W_{j\neq j^{\prime }}$ become small, the
amplitude of QSE jumps decreases reducing, eventually, the saw teeth to
barely visible kinks on $\sigma \left( L\right) $.

If elastic wall scattering is the main scattering mechanism, the usual QSE
oscillations can always be observed for random surface inhomogeneities with
small correlation radius (''size'') $R$, $R<L$. For larger $R$ the interband
transitions are often suppressed making $\sigma \left( L\right) $ smooth,
almost power-law curve. Below we demonstrate that there exists a new type of
QSE oscillations at $R>L$ between two distinct monotonic parts of $\sigma
\left( L\right) $. These new oscillations can be observed only if the
Fourier image $\zeta \left( {\bf q}\right) $ of the correlation function of
random surface inhomogeneities $\zeta \left( {\bf s}\right) $ (the so-called
power spectrum) is rapidly going to zero at large ${\bf q}$. This finding is
illustrated in Figure 1. Curves 1 and 2, which show $\sigma \left( L\right) $%
\ for correlators with exponential power spectra, consist of two smooth
parts separated by an oscillation region. Curves 3 and 4\ for the power-law
spectral functions exhibit usual saw-like QSE. The explanations for the new
QSE and the disappearance of the usual saw-like QSE are interrelated. 
%%%%%%%%%%%%%%%%%%%%%%%%%%%%%%%%%%%%%%%%%%%%%%%%%%%%%%
\begin{figure}[tbp]
%\centerline{\epsfxsize=3.4in\epsfbox{fig1.eps}}
\centerline{\epsfxsize=3.4in\epsfbox{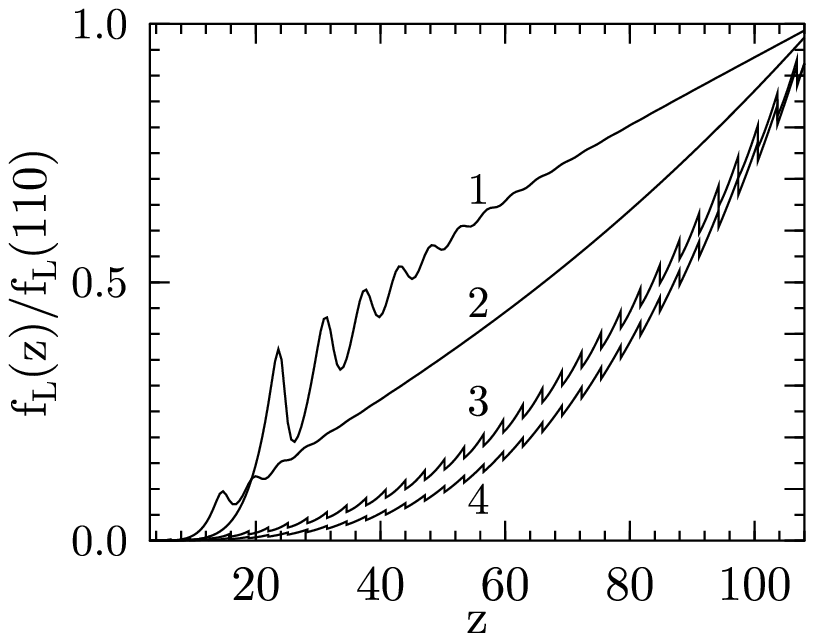}}
\caption{ Normalized functions $f_{L}\left( z\right) $ for $\protect\sigma %
\left( L\right) $, Eq.$\left( \text{\ref{n2}}\right) $, $f_{L}\left(
z\right) /f_{L}\left( z=110\right) $, at $x=200$. Curves 1 and 2
(correlators $\left( \text{\ref{a2}}\right) $\ and $\left( \text{\ref{ee2}}%
\right) $\ with $\protect\mu =0.5$; $f_{L}\left( 110\right) =6.9\cdot
10^{4};7.5\cdot 10^{3}$) exhibit new type of QSE oscillations. Curves 3,4
for surfaces with power spectrum $\left( \text{\ref{ee3}}\right) $ ($\protect%
\lambda =0.5;0$; $f_{L}\left( 110\right) =3.9\cdot 10^{2};1.3\cdot 10^{1}$)
exhibit usual saw-like QSE. }
\label{fig1}
\end{figure}
%%%%%%%%%%%%%%%%%%%%%%%%%%%%%%%%%%%%%%%%%%%%%%
The results are based on the formalism \cite{arm1} that unites earlier
approaches \cite{r3} to transport in systems with random rough walls with or
without bulk scattering. Elastic wall scattering leads to transitions
between the states $\epsilon _{j}\left( {\bf q}\right) \longleftrightarrow
\epsilon _{j^{\prime }}\left( {\bf q}^{\prime }\right) $ with the
probability $W_{jj^{\prime }}\left( {\bf q},{\bf q}^{\prime }\right) $ which
is proportional to the power spectrum of surface inhomogeneities $\zeta
\left( {\bf q}_{j}-{\bf q}_{j^{\prime }}\right) $ ($q_{j}$ is the Fermi
momentum for the miniband $\epsilon _{j}$, $\epsilon _{j}\left( q_{j}\right)
=\epsilon _{F}$). The rate of decrease of $\zeta \left( {\bf q}_{j}-{\bf q}%
_{j^{\prime }}\right) $ at large $q$ depends on the correlation length $R$\
via parameters $\nu _{jj^{\prime }}=R\left| q_{j}-q_{j^{\prime }}\right| $, 
\begin{equation}
\nu _{jj^{\prime }}=\left| \sqrt{z^{2}-\pi ^{2}j^{2}}-\sqrt{z^{2}-\pi
^{2}j^{\prime 2}}\right| R/L  \label{n1}
\end{equation}
where $z=L/\lambda _{F}$. The diagonal $\nu _{jj}=0$. The faster $W_{j\neq
j^{\prime }}$ go to zero with increasing $\nu _{j\neq j^{\prime }}$, the
earlier the transport signs of the usual QSE disappear.

We compared $\sigma \left( L\right) $ for several realistic correlation
functions \cite{q2}: the Gaussian correlator, 
\begin{equation}
\zeta \left( {\bf s}\right) =\ell ^{2}\exp \left( -s^{2}/2R^{2}\right) ,
\label{a2}
\end{equation}
power-law correlators with various $\mu ,$ 
\begin{equation}
\zeta \left( {\bf s}\right) =2\mu \ell ^{2}\left[ R^{2}/\left(
s^{2}+R^{2}\right) \right] ^{1+\mu },  \label{ee2}
\end{equation}
including the Staras correlator $\mu =1$, the Lorentzian 
\begin{equation}
\zeta \left( {\bf s}\right) =2\ell ^{2}R^{2}/\left( s^{2}+R^{2}\right) ,
\label{eee2}
\end{equation}
and the correlators with a power-law Fourier image, 
\begin{equation}
\zeta \left( {\bf q}\right) =2\pi \ell ^{2}\left[ R^{2}/\left(
1+q^{2}R^{2}\right) \right] ^{1+\lambda }.  \label{ee3}
\end{equation}
The last group includes the Lorentzian in momentum space $\lambda =0$ (see
experiment \cite{fer1}) and the exponential correlator $\zeta \left( {\bf s}%
\right) =\ell ^{2}\exp \left( -s/R\right) $ at $\lambda =1/2$. All the
correlators describe the surface inhomogeneities of the same average
amplitude $\ell $ and, except for $\left( \text{\ref{eee2}}\right) $, lead
to the same conductivity $\sigma $ in the long-wave limit $R/\lambda
_{F}\rightarrow 0$ in which $\sigma $ should not depend on details of the
inhomogeneities. The Fourier image of the Lorentzian $\left( \text{\ref{eee2}%
}\right) $ contains the function $K_{0}\left( qR\right) $ and diverges
logarithmically at $R/\lambda _{F}\rightarrow 0$. We do not want to get into
the discussion to what extent this correlator is ''physical''. The fact that
this correlator is used in some calculations \cite{pal1} is sufficient
enough to consider it. To deal with the divergency, one can truncate this
correlator at large distances (commonly, at about 0.1 of the system length 
\cite{q2}). The divergence, by itself, does not lead to any singularities in 
$\sigma $. [Sometimes, the divergence of the power spectrum $\zeta \left( 
{\bf q}\right) $ is associated with a fractal nature of the surface \cite{q2}%
; to what extent our approach can be used for films with fractal surfaces is
an open question].

In all four Figures below curve 1 corresponds to the Gaussian correlator $%
\left( \text{\ref{a2}}\right) $, curve 2 to Eq.$\left( \text{\ref{ee2}}%
\right) $ with $\mu =1/2$, and curves 3 and 4 to Eq.$\left( \text{\ref{eee2}}%
\right) $\ with $\lambda =1/2$ and $0.$

The power spectrum of the Gaussian $\left( \text{\ref{a2}}\right) $\ decays
at large $qR$ as $\exp \left( -q^{2}R^{2}/2\right) $ and the off-diagonal $%
W_{jj^{\prime }}\ $go to zero faster than the diagonal ones by the factor $%
\exp \left( -\nu _{jj^{\prime }}^{2}/2\right) $. The power spectra of the
correlators $\left( \text{\ref{ee2}}\right) ,\left( \text{\ref{eee2}}\right) 
$ contain $\left( qR\right) ^{\mu }K_{\mu }\left( qR\right) $ and $W_{j\neq
j^{\prime }}\ $go to zero as $\nu _{jj^{\prime }}^{\mu -1/2}\exp \left( -\nu
_{jj^{\prime }}\right) $. The slowest, power-law decay of the power spectrum
corresponds to inhomogeneities $\left( \text{\ref{ee3}}\right) $. The
amplitudes of the QSE drops of $\sigma \left( L\right) $ in the points $%
z=L/\lambda _{F}=k\pi $ decrease with increasing $\nu _{kj}$ with the rate
that reflects the dependence $W_{kj}\left( \nu _{kj}\right) $. Accordingly,
the QSE saw disappears, with increasing $R$, first for the surfaces with
Gaussian inhomogeneities, then for the correlators $\left( \text{\ref{ee2}}%
\right) ,\left( \text{\ref{eee2}}\right) $, and almost never for $\left( 
\text{\ref{ee3}}\right) $. This different rate of suppression of QSE is
illustrated in Figure 1 at $x=R/\lambda _{F}=200$. The $2D$ conductivity $%
\sigma \left( L\right) $ is parameterized as 
\begin{equation}
\sigma \left( L\right) =\frac{2e^{2}}{\hbar }\frac{R^{2}}{\ell ^{2}}%
f_{L}\left( z,x\right) .  \label{n2}
\end{equation}
Since $f_{L}\left( z,x=200\right) $ for correlators $\left( \text{\ref{a2}}%
\right) -\left( \text{\ref{ee3}}\right) $ with the same values of $\ell $
and $R$ have different orders of magnitude, functions $f_{L}$, for better
comparison, are normalized by their values at $z=110$, $f_{L}\left( z\right)
/f_{L}\left( 110\right) $. At $x=200$, $\exp \left( -\nu _{j\neq j^{\prime
}}^{2}/2\right) $ and $\exp \left( -\nu _{j\neq j^{\prime }}\right) $ are
small and QSE is suppressed for Gaussian $\left( \text{\ref{a2}}\right) $
and power-law $\left( \text{\ref{ee2}}\right) $ ($\mu =0.5$) correlations
(curves 1,2), but still persists for the slowly decaying power spectra $%
\left( \text{\ref{ee3}}\right) $ with $\lambda =0.5;0$ (curves 3,4).

What is unexpected is the appearance of a new oscillation structure on
curves 1,2 for $z$ between 20 and 90 for the Gaussian and power-law
correlators. It looks as if there are two distinct regimes with large
oscillations in-between. These oscillations are not related to the usual
QSE, {\it i.e.}, to abrupt changes in the number of occupied minibands $%
S\left( z\right) =${\rm Int}$\left( z/\pi \right) $: the oscillations are
less sharp, have a larger period roughly proportional to $z^{2}$, and appear
only at relatively large $z$ and $S$.

The explanation involves the interband transitions. It seems that at large $x
$ the off-diagonal $\nu _{jj^{\prime }}$ $\left( \text{\ref{n1}}\right) $
are large and the interband transitions are suppressed. However, for large $z
$ few of the elements $\nu _{jj^{\prime }}$ with {\it small} $j,$ which are
close to the main diagonal, could become relatively small even for large $x$%
, $\nu _{j,j+1}\left( j\ll z/\pi \right) \sim \pi ^{2}x\left( 2j+1\right)
/2z^{2}$. Then the transitions $j\leftrightarrow j+1$ become noticeable,
\begin{equation}
W_{j,j+1}^{\left( 0\right) }\left( x,z\right) \sim W_{jj}^{\left( 0\right)
}\left( x,z\right) -W_{jj}^{\left( 1\right) }\left( x,z\right)   \label{w1}
\end{equation}
($W_{jj^{\prime }}^{\left( 0,1\right) }$ are the angular harmonics of $%
W\left( {\bf q}_{j}-{\bf q}_{j^{\prime }}\right) $ over the angle $\widehat{%
{\bf q}_{j}{\bf q}_{j^{\prime }}}$). The opening of such transition channels
is accompanied by drops in conductivity. Eqs.$\left( \text{\ref{w1}}\right) $
define the positions $z_{j}\left( x\right) $ of such drops in $\sigma \left(
L\right) $. At $z=z_{1}\left( x\right) $, $W_{12}$ is the first of
transition probabilities to acquire the ''normal'' order of magnitude. At $%
z=z_{2}\left( x\right) $, $W_{23}$ becomes noticeable, then $W_{34}$, {\it %
etc.} The amplitudes of the drops rapidly decrease with increasing $j$. In
the end, when all interband channels with $j\ll z/\pi $ are open, $\sigma
\left( L\right) $ becomes smooth, but with a much lower slope than in its
initial part. The transitions $j\leftrightarrow j+1$ with high $j$ always
remain suppressed at large $x$ and the usual saw-like QSE does not reappear.
The growth of transition probabilities for transitions $j\leftrightarrow j+2$
does not result in new oscillations in $\sigma \left( L\right) $. In the
points $z\left( x\right) $ where $W_{j,j+2}$ becomes large, $%
W_{j,j+2}^{\left( 0\right) }\sim W_{jj}^{\left( 0\right) }-W_{jj}^{\left(
1\right) }$, the states $j$ and $j+2$ are already strongly coupled via $%
W_{j,j+1}$\ and $W_{j+1,j+2}$.

According to \cite{arm1}, for Gaussian inhomogeneities 
\begin{equation}
W_{jj^{\prime }}^{\left( 0,1\right) }=\frac{4\pi ^{5}\ell ^{2}R^{2}}{%
m^{2}L^{6}}\left[ e^{-QQ^{\prime }}I_{0,1}\left( QQ^{\prime }\right) \right]
e^{-\left( Q-Q^{\prime }\right) ^{2}/2},  \label{w2}
\end{equation}
$Q=q_{j}R,$ $Q^{\prime }=q_{j^{\prime }}R$. The asymptotic solution of $%
\left( \text{\ref{w1}}\right) $ is 
\begin{equation}
z_{j}\left( x\right) \approx \frac{\pi }{2}\sqrt{\left( 2j+1\right) x}\left[
\ln \left( x\sqrt{2}\left( 1+1/j\right) \right) \right] ^{-1/4}.  \label{w3}
\end{equation}
The values $z_{j}\left( x=200\right) =24.3;31.7;37.7;....$ agree well with
the positions of the drops on curve 1 of Fig.1.

For the surface with the power-law correlations of inhomogeneities $\left( 
\text{\ref{eee2}}\right) $ with $\mu \lesssim 1$, the solution of Eq.$\left( 
\text{\ref{w1}}\right) $ is not sensitive to $\mu $. With logarithmic
accuracy
\begin{eqnarray}
z_{j}\left( x\right)  &=&\pi \sqrt{\left( 2j+1\right) x/4\nu },  \label{w4}
\\
\nu  &\sim &\ln \left[ x\left( 1+1/i\right) \sqrt{2\ln \left( x\left(
1+1/i\right) \right) }\right] .  \nonumber
\end{eqnarray}

The saw-like drops in conductivity for usual QSE correspond to opening of
transitions to and from the newly accessible, highest miniband while all
other interband transitions are also allowed. The drops are equidistant with
the period $\pi $ along the $z$ axis. The new QSE oscillations in Fig. 1
correspond to the opening of transitions between the lowest minibands while
the transitions in and out of higher minibands are suppressed. The peaks are
almost equidistant if plotted against $z^{2}$.

The initial part of the curves 1,2 for $\sigma \left( L\right) $ is
described analytically by equations of Ref.\cite{arm1} and is close to the
power law $\sigma \propto L^{\left( 5+\alpha \right) }$ (small $\alpha $
depends on $x$) and to experimental data of the first Ref.\cite{qse1}. After
the region of new QSE oscillations, the curves are again smooth, but with a
much smaller tangent. We do not have an analytical description for this
regime. The numerical analysis yields either $\sigma =A+B\cdot L^{1+\beta }$
with small $\beta $ or $a+b\cdot L+c\cdot L^{2}$. This is close to
experimental data \cite{or1} and is different from the known behavior of $%
\sigma \left( L\right) $ at $x=p_{F}R\ll 1$ (see\ second references in \cite
{arm1,r3}). %%%%%%%%%%%%%%%%%%%%%%%%%%%%%%%%%%%%%%%%%%%%%%%%%%%%%%
\begin{figure}[tbp]
\centerline{\epsfxsize=3.4in\epsfbox{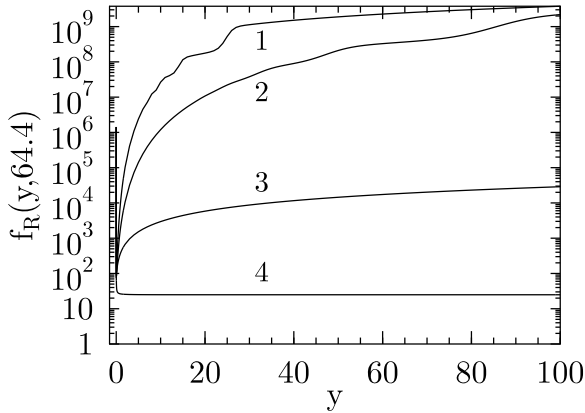}}
%\centerline{\epsfxsize=3.4in\epsfbox{fig2.eps}}
\caption{Functions $f_{R}\left( y\right) $ for $\protect\sigma \left(
R\right) $, Eq.$\left( \text{\ref{c4}}\right) $, at $z=64.4$. Curves 1 and 2
(correlators $\left( \text{\ref{a2}}\right) $\ and $\left( \text{\ref{ee2}}%
\right) $ with $\protect\mu =0.5$) exhibit new QSE (steps). Curves 3,4
(correlators $\left( \text{\ref{ee3}}\right) ,$ $\protect\lambda =0.5;0$)
are smooth in accordance with usual QSE. }
\label{fig2}
\end{figure}
%%%%%%%%%%%%%%%%%%%%%%%%%%%%%%%%%%%%%%%%%%%%%%
The dependence of the conductivity on the correlation radius of surface
inhomogeneities, $\sigma \left( R\right) $, is best illustrated by the
function $f_{R}\left( y,z=const\right) $, 
\begin{equation}
\sigma \left( R\right) =\frac{2e^{2}}{\hbar }\frac{L^{2}}{\ell ^{2}}%
f_{R}\left( y,z=const\right) ,  \label{c4}
\end{equation}
with $y=R/L$. The number of the occupied minibands $S$ does not depend on
the correlation radius of inhomogeneities, \ and $f_{R}\left(
y,z=const\right) $ does not exhibit any saw-like QSE. However, these curves
exhibit the step-like structure that corresponds to the new QSE oscillations
of Fig. 1.

The positions of singularities $y_{j}\left( z\right) $ on $f_{R}\left(
y,z=const\right) $ are identified by Eqs.$\left( \text{\ref{w1}}\right) $
with $x=yz$. The functions $f_{R}\left( y,z=64.4\right) $ are plotted in
Figure 2 for several correlators. The steps on curve 1 in the points $%
y=25;14;8;...$ agree well with the solution $y\left( z\right) $\ of Eq.$%
\left( \text{\ref{w3}}\right) $. The same feature, though barely
discernible, is also observed for the power-law correlators. [Minima in all
curves near the vertical axis describe the region of the most effective
surface scattering at $p_{F}R\sim 1$].

The dependence of the conductivity $\sigma $ on the density of fermions $N$
or their Fermi momentum $p_{F}$ is best displayed by the function $%
f_{N}\left( z\right) $, 
\begin{equation}
\sigma \left( p_{F}\right) =\frac{2e^{2}}{\hbar }\frac{L^{2}}{\ell ^{2}}%
f_{N}\left( z,y=const\right) .  \label{c5}
\end{equation}
Function $\sigma \left( p_{F}\right) $ exhibits usual saw-like QSE at not
very high $y$ for all types of correlators. With increasing $y$, the saw
teeth disappear, first for the Gaussian and later for the power-law
correlators, and persist for the power-law correlators in the momentum
space. %%%%%%%%%%%%%%%%%%%%%%%%%%%%%%%%%%%%%%%%%%%%%%%%%%%%%%
\begin{figure}[tbp]
%\centerline{\epsfxsize=3.4in\epsfbox{fig3.eps}}
\centerline{\epsfxsize=3.4in\epsfbox{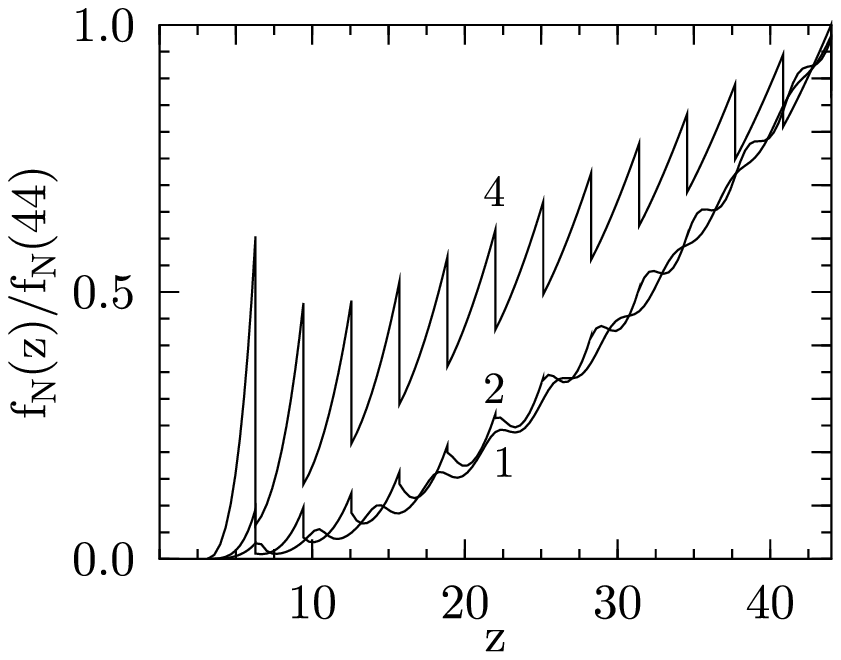}}
\caption{Normalized functions $f_{N}\left( z\right) $ for $\protect\sigma %
\left( p_{F}\right) $, Eq.$\left( \text{\ref{c5}}\right) $, $f_{N}\left(
z\right) /f_{N}\left( z=44\right) $, at $y=R/L=1$. Curves 1 and 2
(correlators $\left( \text{\ref{a2}}\right) $, $\left( \text{\ref{ee2}}%
\right) $ with $\protect\mu =0.5$; $f_{N}\left( 44\right) =5\cdot
10^{3};2.5\cdot 10^{2}$) exhibit suppressed usual QSE peaks at small $z$
that gradually transform at higher $z$ into the new QSE oscillations with
larger period. Curve 4 for surfaces with power spectrum $\left( \text{\ref
{ee3}}\right) $ ($\protect\lambda =0$; $f_{N}\left( 44\right) =20.2$)
exhibits usual QSE. }
\label{fig3}
\end{figure}
%%%%%%%%%%%%%%%%%%%%%%%%%%%%%%%%%%%%%%%%%%%%%%
Curves $f_{N}\left( z\right) $ exhibit the effect related to the new QSE
oscillations of Figure 1 for $\sigma \left( L\right) $ and to the steps in
Figure 2 for $\sigma \left( R\right) $. Figure 3 shows normalized (by the
highest value) functions $f_{N}\left( z\right) $ for the correlators $\left( 
\text{\ref{a2}}\right) $ (curve 1), $\left( \text{\ref{ee2}}\right) $ ($\mu
=1/2,$ curve 2), and $\left( \text{\ref{ee3}}\right) $ ($\lambda =0.5$,
curve 4). The correlation radius $R$ is small, $y=1$, and the figure
illustrates the transition from usual to new QSE. The correlators $\left( 
\text{\ref{ee3}}\right) $\ have a slowly decaying power spectrum and the
functions $f_{N}\left( z\right) $\ reveal usual saw-like QSE. Curve 2 starts
as a usual QSE curve, but, with increasing $z$, the oscillations loose the
saw-like shape and increase the period. Curve 1 for the Gaussian correlator
with a much faster decaying power spectrum does not exhibit, even for the
smallest $z$, neither the shape nor the periodicity of usual QSE.

Curves $f_{N}\left( z\right) $ for the same correlators, but at $y=20$, are
shown in Figure 4. Curves 3,4 still exhibit usual QSE, while curves 1,2 show
the well-developed oscillations of the new type. The peaks on curve 1 at $%
z_{j}\left( y=20\right) =19.8;50.3;83.6;...$ are in good agreement with Eq.$%
\left( \text{\ref{w3}}\right) $ with $x=yz$. 
%%%%%%%%%%%%%%%%%%%%%%%%%%%%%%%%%%%%%%%%%%%%%
\begin{figure}[tbp]
%\centerline{\epsfxsize=3.4in\epsfbox{fig4.eps}}
\centerline{\epsfxsize=3.4in\epsfbox{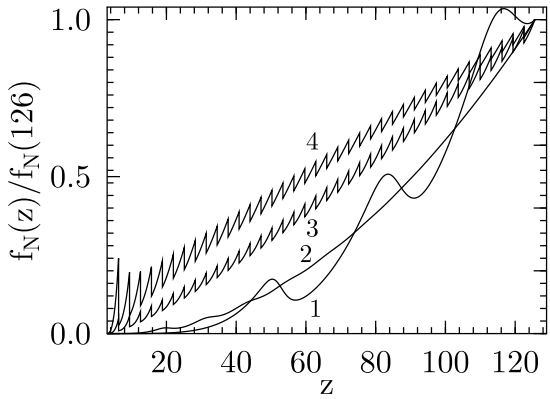}}
\caption{$f_{N}\left( z\right) /f_{N}\left( z=126\right) $ for $y=R/L=20$.
Curves 1,2 (correlators $\left( \text{\ref{a2}}\right) $ and $\left( \text{%
\ref{ee2}}\right) $, $\protect\mu =0.5$; $f_{N}\left( 126\right) =1.1\cdot
10^{9};4.5\cdot 10^{7}$) exhibit well-developed new QSE oscillations. Curves
3,4 \ for correlators $\left( \text{\ref{ee3}}\right) $ with $\protect%
\lambda =0.5;0$ ($f_{N}\left( 126\right) =1.4\cdot 10^{4};47$) still exhibit
usual saw-like QSE. }
\label{fig4}
\end{figure}
%%%%%%%%%%%%%%%%%%%%%%%%%%%%%%%%%%%%%%%%%%%%%%
In summary, we predict new type of QSE in conductivity of films with random
rough walls. The effect is reminiscent of magnetic breakthrough. The
positions of the peaks $\left( \text{\ref{w1}}\right) $ are determined by
the angular harmonics of the correlation function of surface
inhomogeneities. These new QSE singularities replace usual QSE for surface
inhomogeneities with large correlation radius and with rapidly
(exponentially) decaying power spectra such as for Gaussian or power-law
correlation functions. Surfaces with the power-law decay of the Fourier
image of the correlation functions exhibit persistent standard QSE and do
not exhibit new singularities. Dependences of the conductivity on the film
thickness, correlation radius of inhomogeneities, and the particle density
(Fermi momentum) display these new QSE anomalies in consistent, but somewhat
different, ways. Analysis of transport along surfaces with large $R$ by
equations for usual QSE can result in misinterpretation of experimental
data. Large period of new oscillations can make the observation of QSE in
metal films easier ({\it cf.} the last of Refs.\cite{qse1}).

The work is supported by NSF grant DMR-0077266.

\end{document}